\documentclass{pasj00}
\begin{document}
\SetRunningHead{Lyratzi et al.}{The complex structure of the Mg II
region of 64 Be stars}
\Received{2000/12/31}%{yyyy/mm/dd}
\Accepted{2001/01/01}%{yyyy/mm/dd}

\title{The complex structure of the Mg II $\lambda\lambda$ 2795.523, 2802.698 ${\rm\AA}$ regions of 64 Be stars}

%%% begin:list of authors
\author{Evaggelia \textsc{Lyratzi}%
%\thanks{Example: Present Address is xxxxxxxxxx}
}
\affil{University of Athens, Faculty of Physics Department of
        Astrophysics, Astronomy and Mechanics}
\affil{Panepistimioupoli, Zographou 157 84, Athens, Greece}
\email{elyratzi@phys.uoa.gr}

\author{Emmanouel \textsc{Danezis}}
\affil{University of Athens, Faculty of Physics Department of
        Astrophysics, Astronomy and Mechanics}
\affil{Panepistimioupoli, Zographou 157 84, Athens, Greece}
\email{edanezis@phys.uoa.gr}

\author{Luka \v{C}. \textsc{Popovi\'{c}}}%
\affil{Astronomical Observatory of Belgrade, Volgina 7, 11160
        Belgrade, Serbia}
\affil{Isaac Newton Institute of Chile, Yugoslavia
        Branch}
\email{lpopovic@aob.bg.ac.yu}

\author{Milan S. \textsc{Dimitrijevi\'{c}}}%
\affil{Astronomical Observatory of Belgrade, Volgina 7, 11160
        Belgrade, Serbia}
\affil{Isaac Newton Institute of Chile, Yugoslavia
        Branch}
\email{mdimitrijevic@aob.bg.ac.yu}

\author{Dimitris \textsc{Nikolaidis}}
\affil{University of Athens, Faculty of Physics Department of
        Astrophysics, Astronomy and Mechanics}
\affil{Panepistimioupoli, Zographou 157 84, Athens, Greece}

\and
\author{Antonis {\sc Antoniou}}
\affil{University of Athens, Faculty of Physics Department of
        Astrophysics, Astronomy and Mechanics}
\affil{Panepistimioupoli, Zographou 157 84, Athens, Greece}
\email{ananton@phys.uoa.gr}
%%% end:list of authors

%%% Please use the following style in case that sorting by
%%% affilation is impossible.
%
% \author{%
%   Evaggelia \textsc{Lyratzi}\altaffilmark{1}
%   Emmanouel \textsc{Danezis}\altaffilmark{1}
%   Luka \v{C}. \textsc{Popovi\'{c}}\altaffilmark{2,3}
%   Milan S. \textsc{Dimitrijevi\'{c}}\altaffilmark{2,3}
%   Dimitris \textsc{Nikolaidis}\altaffilmark{1}
%   and
%   Antonis \textsc{Antoniou}\altaffilmark{1}}
% \altaffiltext{1}{University of Athens, Faculty of Physics Department of
%        Astrophysics, Astronomy and Mechanics Panepistimioupoli,
%        Zographou 157 84, Athens, Greece}
% \email{elyratzi@phys.uoa.gr}
% \altaffiltext{2}{Astronomical Observatory of Belgrade, Volgina 7, 11160
%        Belgrade, Serbia}
% \altaffiltext{3}{Isaac Newton Institute of Chile, Yugoslavia
%        Branch}

%% `\KeyWords{}' always has to be placed before `\maketitle'.
%Do NOT move this preamble from here!

\maketitle

\begin{abstract}
Here is studied the presence of absorption components shifted to
the violet or the red side of the main spectral line (satellite,
or discrete absorption components, i.e. SACs or DACs), in Mg II
resonance lines' regions in Be stars and their kinematical
characteristics. Namely our objective is to check if exists a
common physical structure for the atmospheric regions creating
SACs or DACs of the Mg II resonance lines. In order to do this, a
statistical study of the Mg II $\lambda \lambda$ 2792.523,
2802.698 $\AA$ lines in the spectra of 64 Be stars of all spectral
subtypes and luminosity classes is performed. We found that the
absorption atmospherical regions where the Mg II resonance lines
originated may be formed of several independent density layers of
matter which rotate with different velocities. It is attempted
also to separate SACs and DACs according to low or high radial
velocity. The emission lines were detected only in the earliest
and latest spectral subtypes.
\end{abstract}

\section{Introduction}

The Mg II resonance lines have a peculiar profile in the Be
stellar spectra which indicates multicomponent nature of their
origin region. Many researchers have observed the existence of
absorption components shifted to the violet or the red side of the
main spectral line \citep{und70, mar78, dac80, doa82, dan84,
dan87, dan91, sah84, sah85, hut85, doa91, las92, cid98, lyr03}.
These components, so called Discrete Absorption Components -- DACs
\citep{bat86} or Satellite Absorption Components -- SACs
\citep{dan03, lyr04}, probably originate in separated regions
which have different rotational and radial velocities. Especially
in the case of very narrow DACs and SACs, they  cannot  be
photospheric, but rather they have circumstellar or interstellar
origin \citep{sle78}. For example, \citet{kon76} found that the
shell absorption gets stronger from intermediate to the late B
stars and suggested that ``it might be due to the rising
temperature in the gaseous shell which converts Mg II to Mg III
and to the weakening of the outward-driving mechanisms of the
atmosphere''. In any case, the whole feature of the Mg II
resonance lines is not the result of a uniform atmospherical
region, but the components are created in different regions, which
rotate and move radially with different velocities. As
\citet{dej79} proposed in their study of 33 stellar spectra of all
the spectral types, in the late B supergiants variable mass loss
occurs, due to ``occasional stellar ``puffs'' superposed on a more
or less regular wind''. They proposed that ``there are
concentrations of low-ionization species in the stellar wind as a
result of the occurrence of significant density variations''.
Also, in order to explain the complex profiles of the Mg II
resonance lines, \citet{cid98} proposed that the Be stellar
atmospheres are composed from a classical photosphere, an
extending high temperature chromosphere and a cool envelope. There
is a question about contribution of different atmospherical
layers, especially in the case of SACs and/or DACs phenomena, in
the construction of the Mg II resonance lines' profiles.

The aim of this work is to statistically investigate the presence
of SACs and/or DACs in Mg II resonance lines' regions in Be stars
and their kinematical characteristics. Also, we would like to
conclude about the limits of the rotational and radial velocities
($V_{rot}$, $V_{rad}$), as well as to check whether there exists a
common physical structure for the atmospherical regions which
create the Satellite Absorption Components (SACs) of the Mg II
resonance lines in the spectra of all the Be stars. To do that, a
statistical study of the UV Mg II resonance lines $\lambda\lambda$
2795.523, 2802.698 \AA\  in the spectra of 64 Be stars (all the
spectral subtypes and luminosity classes) has been performed. The
study is based on the model proposed by \citet{dan03} and
\citet{lyr04}.

We decided to study the Mg II resonance lines, as they are
characteristic of the cool envelope in Be stellar atmospheres,
they are very intense features in the spectra of Be stars and they
mostly present a complex and peculiar structure. Besides, since
they are resonance lines, they give us a possibility to test the
validity of the proposed model, as we have to adhere to all the
necessary physical criteria and techniques \citep{dan03}. Our
purpose here was to study spectral lines which are created in cool
regions (Mg II) in Be stellar atmospheres. \citet{dan87, sah84,
sah85} detected multistructure in the regions where Fe II lines
(I.P.=7.870 eV) are created in the spectra of Be stars which are
characterized as iron stars. One of our purposes was to
investigate whether the multistructure appears only in the Fe II
spectral lines or in other spectral lines also with similar
Ionization Potential, as Mg II (I.P.=7.646 eV). This is another
reason why we chose to study the Mg II resonance lines and not
some other resonance lines of the cool envelope, as N II
(I.P.=14.490 eV), C II (I.P.=11.260 eV), Si II (I.P.=8.110 eV) or
Al II (I.P.=5.986 eV). The study of Mg II resonance lines, gives
us information not only about the cool envelope, but also for the
multistructure of another ion that lies in the same region where
the Fe II spectral lines are created.

In \S 2 we describe the method of analysis, in \S 3 the
observational data, in \S 4 the results and their discussion is
given and in \S 5 we give our conclusions.

\section{Method of spectral line analysis}

Before starting with the description of the method used in our
study, let us explain the differences between DACs and SACs. The
DACs are components of a spectral line of a specific ion, shifted
at different $\Delta\lambda$ from the transition wavelength of a
line, because they are created in  different density regions which
rotate and move radially with different velocities \citep{lyr04}.
The DACs are discrete lines, easily observed, in the spectra of
some Be stars of luminosity class III, in the case of the Mg II
doublet. However, if the layers that give rise to such lines
rotate with quite large velocities and move radially with small
velocities, then the produced lines are quite broadened and
slightly shifted. As a result, they are blended among themselves
as well as with the main spectral component and thus they do not
appear as discrete, consequently they cannot be resolved. In such
a case the name Discrete Absorption Component is inappropriate.
Besides, as \citet{pet74} first pointed out, these components
appear as ``satellites'' in the violet or in the red side of the
main spectral line's component as a function of the time or the
phase in the case of a binary system. For these two reasons and in
order to include all these lines that are components of the line
profile either they are discrete or not to a unique name, it has
been proposed \citep{lyr04} that they should be named Satellite
Absorption Components (SACs) as a general expression, because the
Discrete Absorption Components are ``Satellites'' of a main
spectral line, while the Satellite Absorption Components are not
always discrete and cannot be easily resolved.

An additional peculiar phenomenon is that not all the lines of a
specific ion have DACs. The DACs phenomenon is not a general one,
but it is present in the case of some lines which have low
excitation potential. For example, while the Mg II resonance lines
at 2795.523, 2802.698 \AA\  (multiplet 1) have DACs in the spectra
of some Be stars, their subordinate lines at 2790.768, 2797.989
\AA\ (multiplet 3) do not present the same phenomenon. Therefore,
we decided to study the Mg II resonance lines, as they are the
only Mg II doublet having DACs.

\begin{figure}
  \begin{center}
    \FigureFile(85mm,38mm){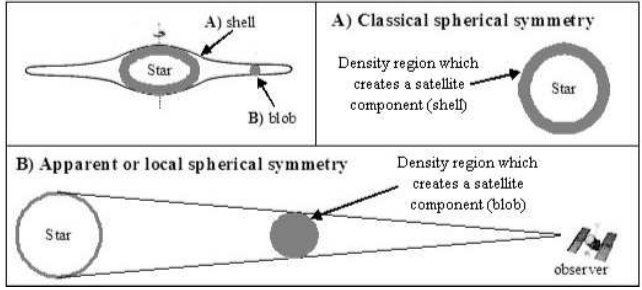}
  \end{center}
  \caption{Density regions which create the observed SACs or DACs in
the stellar spectra.}\label{geo1}
\end{figure}

In order to try to obtain a qualitative picture we will use as an
approximation the spherical geometry. According to it, there are
two possibilities (see Fig. ~\ref{geo1}).

\noindent A) The region which creates the SACs or DACs may be an
envelope around and near the rapidly rotating star. In such a
case, the spherical model could give only a rough approximation
but allow however to obtain some qualitative results. We note that
such assumption is valid for some earlier models like e.g. Doazan
and Thomas (1982) which can not explain the observed free-free
emission or polarization, but we are interested here only in
kinematic characteristics. In this case the calculated values of
the radial velocity correspond to the component of the expansion
or contraction velocity of the shell, which is projected to the
observational axis.

\begin{figure}
  \begin{center}
    \FigureFile(85mm,54mm){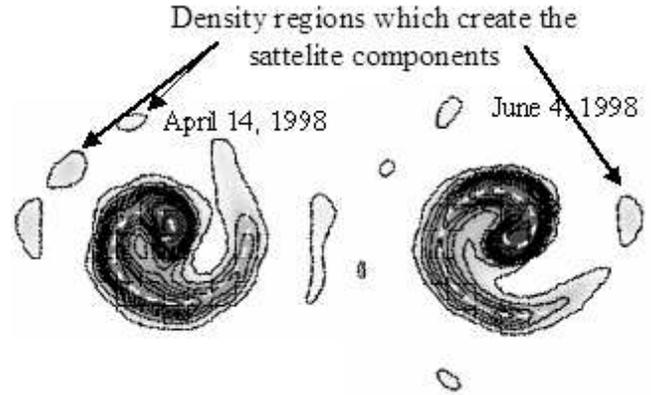}
  \end{center}
  \caption{Due to the rotation of the star the ejected matter forms
spiral streams which produce blobs that create the observed SACs
or DACs in the stellar spectra. This figure is taken from
\citet{tut99}} \label{geo2}
\end{figure}

\noindent B) The region which creates the SACs or DACs may be an
independent density region (blob), which is spherically symmetric
around its own center. Such density regions have been observed
around stars that eject mass. In (Fig.~\ref{geo2}) we see this
phenomenon in WR 104, observed by \citet{tut99}. Such a region may
be the one, which creates the Mg II resonance lines and which lies
in the cool envelope of the stellar atmosphere. Such a blob may
have three different motions: a) it may rotate around the star, b)
it may expand or contract and c) it may move radially. This means
that the calculated values of the radial velocity consist of three
different components: a) the component of the rotational velocity
of the blob around the star, projected to the observational axis,
b) the component of the expansion or contraction velocity of the
blob, projected to the observational axis and c) the component of
the velocity of the blob's radial motion, projected to the line of
sight.

In principle, as it is known, the star ejects mass with a specific
radial velocity. The stream of matter is twisted, forming density
regions such as interaction of fast and slow wind components,
corrotating interaction regions (CIRs), structures due to magnetic
fields or spiral streams as a result of the stellar rotation
\citep{und84, mul84a, mul84b, mul86, pri88, cra96, ful97, kap96,
kap97, kap99, cra00}. Consequently, hydrodynamic and magnetic
forces act as centripetal forces, resulting that the outward
moving matter twists and moves around the star (see
Fig.~\ref{geo2}) . This motion of the matter is responsible for
the formation of high density regions (shells, blobs, puffs,
spiral streams), which, either they are spherically symmetric with
respect to the star or with respect to their own center
\citep{dan03, lyr04} (see Fig.~\ref{geo1}).

In order to study the physical structure and the existence of SACs
phenomena in the regions where these lines are created we used the
model proposed by \citet{dan03} and \citet{lyr04}. This model
allows us to calculate the rotational ($V_{rot}$) and radial
velocities ($V_{rad}$) of independent density layers of matter in
these regions, as well as the optical depth ($\tau$) and the
column density (n). If the considered SACs or DACs originate from
"pufs" or "blobs" created by stellar winds in a cool extended
envelope, the real shape of the envelope is not crucial. If the
considered features are created in the envelope's layers we also
could obtain  some qualitative conclusions, since we are
interested only in kinematical properties here. Moreover, there is
no a sophisticated or rough model of Be star which could fit the
observed profiles considering their complex structure so that
there is no results of similar investigations for comparisons.

Let us consider that the area of gas, which creates a specific
spectral line, consists of $i$ independent absorbing density
regions followed by $j$ independent regions that both absorb and
emit and an outer absorbing region. Such a structure produces DACs
or SACs in the observed spectra \citep{dan03} and the final line
function which can describe the complex line profiles of the
observed spectral lines is \citep{dan03, lyr04}:

\begin{displaymath}
F_{\lambda}=[F_{\lambda0}\prod_{i}
e^{-L_{i}\xi_{i}}+\sum_{j}S_{\lambda
ej}(1-e^{-L_{ej}\xi_{ej}})]e^{-L_{g}\xi_{g}}
\end{displaymath}
where $F_{\lambda}$ is the observed flux, $F_{\lambda0}$ is the
initial flux, $S_{\lambda ej}$ are the source functions of each
emitting  density region, $\xi$ is the optical depth in the center
of the line and $L_{i}$, $L_{ej}$, $L_{g}$ are the distribution
functions of the absorption coefficients $k_{\lambda i}$,
$k_{\lambda ej}$, $k_{\lambda g}$, respectively. Each $L$ depends
on the values of the rotational and the radial velocity of the
density region, which forms each component of the spectral line
($V_{rot}$, $V_{rad}$) \citep{dan03, lyr04}. The product of $L$
and $\xi$ is the optical depth of each region.

This function does not depend on the geometry of the regions
creating the observed feature. The considered geometry of the
regions is taken into account in order to define the distribution
function $L$. This means that $L$ may represent any distribution
which considers certain geometry (see Appendix A), without
changing anything in $F_{\lambda}$.

Each component of the spectral line, which is formed by the
$i^{th}$ density region of matter, must be accurately reproduced
by the function $e^{ - L_{i} \xi _{i} }$ by applying the
appropriate values of $V_{rot_{i}}$, $V_{rad_{i}}$ and $\xi_{i}$.
Using the best model's fit for a complex spectral line, we can
calculate the apparent radial and rotational velocity
($V_{rad_{i}}, V_{rot_{i}}$) and the optical depth ($\xi_{i}$) in
the center of the line of the region in which the main spectral
line and its SACs are created.

In the case that we want to consider that some other physical
parameters are responsible for the line broadening and not the
rotation of the region which produces the studied spectral lines,
we may replace the exponential $e^{-L\xi}$ with another classical
distribution.

In the case of emission lines, each emission component, which is
formed by the $j^{th}$ density region of matter, must be
accurately reproduced by the function $S_{\lambda
ej}(1-e^{-L_{ej}\xi_{ej}})$, by applying the appropriate values of
$V_{rot_{i}}$, $V_{rad_{i}}$, $\xi_{i}$ and $S$. Using the best
model's fit for a complex spectral line, we can calculate the
apparent radial and rotational velocity ($V_{rad_{i}},
V_{rot_{i}}$), the optical depth ($\xi_{i}$) in the center of the
line and the source function $S$ of the region in which the
emission component is created.

The proposed model is relatively simple, aiming to describe the
regions where the spectral lines which present SACs are created.
With this model we study the regions of a specific ion which
creates a specific spectral line.

This model presupposes that the main reason of the line broadening
is the rotation of the region that gives rise to the spectral line
\citep{doa82}. We can accept this assumption when we deal with the
inner layers to the post-coronal regions. Thus, for these
atmospheric layers the model gives satisfactory results.

\section{Observational data and fitting procedure}

The data we used are the Mg II resonance lines of 64 Be stars
taken from the IUE Archive Search
database\footnote{http://archive.stsci.edu/cgi-bin/iue}. The
stellar spectra were observed with IUE satellite using the Long
Wavelength range Prime and Redundant cameras (LWP, LWR) at high
resolution (0.1 to 0.3 \AA). In Table ~\ref{data} we give the list
of stars, their spectral type and the type of camera used during
observations.

Our first step is to identify the spectral lines in the studied
wavelength range, in order to find out which lines may be blended
with the Mg II doublet and, thus, may contribute to the observed
features. The identification has been made by NIST Atomic Spectra
Database\footnote{http://physics.nist.gov/cgi-bin/AtData/lines\_form},
as well as the catalogues of \citet{moo68} and \citet{kel79}. In
this specific spectral range the adjacent features of the Mg II
profiles are intense, but they are away from the Mg II spectral
features, so that, in spite of  their important influence on the
wings, their much smaller influence in central parts is not
important for our discussion. Moreover, as we deal with resonance
lines, we know that if one line of the doublet is well fitted, we
should apply the same parameters to the other one, even if the fit
is not so good. In this case the unfitted regions correspond to
blends. Also, the level of continuum is calculated for the whole
spectrum taken by IUE. This means that its suppression by the
blends does not affect our calculations.

The procedure that we followed for the decomposition of the lines
is described in \citet{dan03}. The way, the criteria and
discussion about the model we use to fit the observed spectral
lines are presented in \citet{dan03} and they are explained in
more detail in \citet{lyr04} and an application of the method is
given in \citet{pop04}. This means that we tried to fit the
observed profiles of the Mg II resonance lines with the less
possible components. At first we tried to fit the observed
spectral feature with one component. If this had not been
possible, we added one more component and then another, until we
accomplished the best fit with the less possible components. This
means that we fitted the observed Mg II features with one to four
components, for different stars. Our results are shown in Tables
~\ref{absrp} and ~\ref{emission}.

\onecolumn

\begin{table}
  \caption{The list of Be stars with spectral type (columns 2, 5)
and the type of the camera used during the observations (columns
3, 6).} \label{data}
  \begin{center}
    \begin{tabular}{lllclllc}
\hline
Star &Spectral Type &Camera &ref &\hspace{1cm}Star &Spectral Type &Camera &ref  \\
\hline
HD 5394         &B0 IV : evar   &Lwr 07861  & 1   &\hspace{0.5cm}HD 25940       &B3 V e     &Lwr 05950  &2\\
HD 53367\**     &B0 IV : e      &Lwr 09286  & 1   &\hspace{0.5cm}HD 45725\**    &B3 V e     &Lwp 10041  &2\\
HD 203374\***   &B0 IV pe       &Lwp 07400  & 1   &\hspace{0.5cm}HD 183362      &B3 V e     &Lwp 11044  &2\\
HD 206773       &B0 V : pe      &Lwr 14808  & 1   &\hspace{0.5cm}HD 208057      &B3 V e     &Lwp 29221  &2\\
HD 200310       &B1 V e         &Lwr 09544  & 1   &\hspace{0.5cm}HD 205637      &B3 V : p   &Lwr 05947  &4\\
HD 212571       &B1 V e         &Lwr 05948  & 2   &\hspace{0.5cm}HD 217543      &B3 V pe    &Lwp 13326  &2\\
HD 44458\**     &B1 V pe        &Lwp 30173  & 1   &\hspace{0.5cm}HD 217050\***  &B4 III pe  &Lwr 05933  &2\\
HD 200120       &B1.5 V nne     &Lwr 11035  & 2   &\hspace{0.5cm}HD 89884       &B5 III     &Lwp 29529  &6\\
HD 193237       &B2 pe          &Lwr 07990  & 2   &\hspace{0.5cm}HD 22192       &B5 V e     &Lwr 06898  &2\\
HD 45910\**     &B2 III e       &Lwr 12138  & 3   &\hspace{0.5cm}HD 23302       &B6 III e   &Lwr 09071  &2\\
HD 37202        &B2 IV p        &Lwr 05888  & 4   &\hspace{0.5cm}HD 45542       &B6 III e   &Lwp 07631  &2\\
HD 36576        &B2 IV - V e    &Lwp 14029  & 2   &\hspace{0.5cm}HD 109387      &B6 III pe  &Lwr 04132  &2\\
HD 212076       &B2 IV - V e    &Lwr 03406  & 2   &\hspace{0.5cm}HD 23480       &B6 IV e    &Lwr 05219  &2\\
HD 32991\**     &B2 V e         &Lwr 11426  & 2   &\hspace{0.5cm}HD 217891      &B6 V e     &Lwr 09069  &2\\
HD 58050        &B2 V e         &Lwr 14810  & 2   &\hspace{0.5cm}HD 138749      &B6 V nne   &Lwr 07858  &2\\
HD 164284       &B2 V e         &Lwr 11038  & 2   &\hspace{0.5cm}HD 23630       &B7 III     &Lwr 09060  &2\\
HD 41335        &B2 V ne        &Lwr 07384  & 2   &\hspace{0.5cm}HD 209409      &B7 IV e    &Lwp 15464  &2\\
HD 52721        &B2 V ne        &Lwp 05462  & 5   &\hspace{0.5cm}HD 6811        &B7 V e     &Lwr 09070  &7\\
HD 58343        &B2 V ne        &Lwr 07363  & 6   &\hspace{0.5cm}HD 192044\**   &B7 V e     &Lwp 08135  &8\\
HD 148184       &B2 V ne        &Lwr 06744  & 6   &\hspace{0.5cm}HD 210129      &B7 V ne    &Lwp 23173  &7\\
HD 202904       &B2 V e         &Lwr 07343  & 2   &\hspace{0.5cm}HD 142983      &B8 Ia/Iab  &Lwr 07359  &6\\
HD 65079        &B2 V ne        &Lwp 30119  & 5   &\hspace{0.5cm}HD 29866       &B8 IV ne   &Lwr 08745  &2\\
HD 28497        &B2 V ne        &Lwr 07337  & 6   &\hspace{0.5cm}HD 47054       &B8 V e     &Lwp 13074  &9\\
HD 45995\**     &B2 V nne       &Lwr 08648  & 1   &\hspace{0.5cm}HD 183914\**   &B8 V e     &Lwr 04609  &9\\
HD 10516        &B2 V pe        &Lwr 07335  & 2   &\hspace{0.5cm}HD 50138       &B8 V e     &Lwr 09358  &11\\
HD 187567\**    &B2.5 IV e      &Lwp 14025  & 2   &\hspace{0.5cm}HD 58715       &B8 V var   &Lwp 10104  &10\\
HD 191610       &B2.5 V e       &Lwr 07342  & 2   &\hspace{0.5cm}HD 23552\***   &B8 V ne    &Lwr 08744  &7\\
HD 32343\**     &B2.5 V e       &Lwr 05890  & 2   &\hspace{0.5cm}HD 185037      &B8 V ne    &Lwp 08136  &9\\
HD 65875        &B2.5 V e       &Lwr 05616  & 2   &\hspace{0.5cm}HD 199218      &B8 V nne   &Lwp 09903  &7\\
HD 60855\***    &B2/B3 V        &Lwp 15477  & 6   &\hspace{0.5cm}HD 91120       &B8/B9 IV/V &Lwp 07475  &6\\
HD 37490\**     &B3 III e       &Lwr 07361  & 1   &\hspace{0.5cm}HD 142926      &B9 pe      &Lwr 05768  &7\\
HD 50820        &B3 IV e+...    &Lwr 16776  & 2   &\hspace{0.5cm}HD 144         &B9 III e   &Lwr 08997  &7\\
\hline
    \end{tabular}
  \end{center}
\** Double system \*** Triple system\\
References: (1) Morgan, Code \& Whitford 1955; (2) Lesh 1968; (3)
Sahade, Brandi \& Fontela 1984; (4) Herbig \& Spalding 1955; (5)
Guetter 1968; (6) Houk N. \& Smith-Moore 1988; (7) Cowley 1972;
(8) Osawa 1959; (9) Cowley et al. 1969; (10) Slettebak 1954; (11)
Houziaux
\& Andrillat 1976.\\
\end{table}

\begin{longtable}{llrrrrrrrr}
\caption{The kinematical parameters for absorption regions for
considered stars: the rotational ($V_{rot}$) and the radial
($V_{rad}$) velocities in $\rm {km\ s^{-1}}$.}
\label{absrp}
\endhead
\hline
Star & Spectral Type & $V_{rot1}$ & $V_{rad1}$ & $V_{rot2}$ & $V_{rad2}$ & $V_{rot3}$ & $V_{rad3}$ & $V_{rot4}$ & $V_{rad4}$  \\
\hline
HD 5394     &B0 IV evar     &25         &$8.8$      &50         &$12.3$         &           &               &           &       \\
HD 53367    &B0 IV : e      &24         &$-18.0$    &37         &$-12.0$        &7l         &$-7.0$         &           &       \\
HD 203374   &B0 IV pe       &24         &$4.5$      &36         &$3.5$          &60         &$2.0$          &           &       \\
HD 206773   &B0 V : pe      &28         &$15.5$     &46         &$18.5$         &           &               &           &       \\
HD 200310   &B1 V e         &25         &$10.1$     &           &               &           &               &           &       \\
HD 212571   &B1 V e         &22         &$-2.5$     &40         &$-11.0$        &           &               &           &       \\
HD 44458    &B1 V pe        &22         &$-17.5$    &37         &$-19.5$        &83         &$-29.0$        &           &       \\
HD 200120   &B1.5 V nne     &28         &$-0.4$     &           &               &           &               &           &       \\
HD 193237   &B2 pe          &30         &$-18.6$    &45         &$-18.1$        &86         &(DAC) $-200.6$ &115        &(DAC) $-200.1$\\
HD 45910    &B2 III e       &24         &$13.0$     &42         &$11.0$         &60         &(DAC) $-229.5$ &170        &(DAC) $-182.0$\\
HD 37202    &B2 IV p        &29         &$-18.0$    &48         &$-21.0$        &           &               &125        &$-33.0$\\
HD 36576    &B2 IV - V e    &29         &$-43.0$    &           &               &80         &$-21.0$        &           &       \\
HD 212076   &B2 IV - V e    &20         &$-1.0$     &34         &$-1.5$         &           &               &           &       \\
HD 32991    &B2 V e         &19         &$-15.4$    &38         &$-21.9$        &           &               &           &       \\
HD 58050    &B2 V e         &25         &$-35.9$    &49         &$-28.4$        &           &               &           &       \\
HD 164284   &B2 V e         &28         &$13.3$     &           &               &63         &$33.8$         &           &       \\
HD 41335    &B2 V ne        &22         &$-9.5$     &38         &$-1.5$         &80         &$27.0$         &100        &$40.0$\\
HD 52721    &B2 V ne        &18         &$12.3$     &34         &$11.3$         &65         &$31.3$         &           &       \\
HD 58343    &B2 V ne        &21         &$-1.8$     &50         &$-5.2$         &           &               &           &       \\
HD 148184   &B2 V ne        &17         &$19.9$     &30         &$20.9$         &           &               &           &       \\
HD 202904   &B2 V e         &20         &$4.0$      &31         &$4.5$          &           &               &           &       \\
HD 65079    &B2 V ne        &29         &$8.0$      &           &               &74         &$-12.5$        &           &       \\
HD 28497    &B2 V ne        &8          &$-45.0$    &37         &$-34.5$        &60         &$-35.0$        &           &       \\
HD 45995    &B2 V nne       &29         &$-17.0$    &45         &$-20.0$        &69         &$-24.0$        &           &       \\
HD 10516    &B2 V pe        &22         &$-1.0$     &37         &$-2.3$         &50         &$-5.8$         &           &       \\
HD 187567   &B2.5 IV e      &19         &$23.6$     &32         &$24.6$         &50         &$26.1$         &           &       \\
HD 32343    &B2.5 V e       &20         &$8.4$      &           &               &78         &$17.9$         &           &       \\
HD 65875    &B2.5 V e       &17         &$14.0$     &35         &$18.0$         &           &               &           &       \\
HD 191610   &B2.5 V e       &20         &$-45.6$    &45         &$-53.6$        &           &               &           &       \\
HD 60855    &B2/B3 V        &           &           &45         &$-14.6$        &71         &$22.9$         &139        &$-29.6$\\
HD 37490    &B3 III e       &20         &$-5.3$     &35         &$-8.3$         &54         &$-18.8$        &98         &$-10.3$\\
HD 50820    &B3 IV e+...    &24         &$-13.0$    &50         &$-9.5$         &           &               &           &       \\
HD 25940    &B3 V e         &22         &$0.7$      &42         &$-2.3$         &           &               &           &       \\
HD 45725    &B3 V e         &18         &$-9.5$     &34         &$-10.5$        &65         &$-13.0$        &           &       \\
HD 183362   &B3 V e         &24         &$24.2$     &34         &$24.2$         &68         &$19.7$         &           &       \\
HD 208057   &B3 V e         &23         &$12.3$     &           &               &75         &$9.8$          &           &       \\
HD 205637   &B3 V : p       &22         &$17.5$     &39         &$15.5$         &           &               &115        &$9.0$\\
HD 217543   &B3 V pe        &26         &$16.5$     &51         &$15.5$         &           &               &130        &$23.5$\\
HD 217050   &B4 III pe      &35         &$14.3$     &54         &$17.8$         &84         &$32.3$         &           &       \\
HD 89884    &B5 III         &           &           &33         &$-9.5$         &60         &$1.5$          &           &       \\
HD 22192    &B5 V e         &16         &$-2.0$     &32         &$-1.0$         &52         &$-7.0$         &           &       \\
HD 23302    &B6 III e       &21         &$-9.9$     &           &               &86         &$-8.9$         &           &       \\
HD 45542    &B6 III e       &15         &$-24.4$    &33         &$-24.9$        &58         &$-15.4$        &           &       \\
HD 109387   &B6 III pe      &18         &$-11.4$    &           &               &66         &$-28.4$        &           &       \\
HD 23480    &B6 IV e        &20         &$-7.7$     &32         &$0.3$          &           &               &           &       \\
HD 217891   &B6 V e         &18         &$-0.8$     &           &               &           &               &           &       \\
HD 138749   &B6 V nne       &27         &$22.5$     &47         &$14.5$         &           &               &           &       \\
HD 23630    &B7 III         &20         &$-9.0$     &45         &$-9.0$         &           &               &           &       \\
HD 209409   &B7 IV e        &21         &$-9.0$     &38         &$-9.0$         &83         &$-5.0$         &           &       \\
HD 6811     &B7 V e         &19         &$3.3$      &35         &$1.8$          &           &               &169        &$12.8$\\
HD 192044   &B7 V e         &25         &$19.5$     &52         &$8.5$          &           &               &           &       \\
HD 210129   &B7 V ne        &21         &$62.6$     &52         &$43.1$         &           &               &           &       \\
HD 142983   &B8 Ia/Iab      &21         &$-7.0$     &39         &$8.0$          &60         &$11.5$         &95         &$16.5$\\
HD 29866    &B8 IV ne       &26         &$-42.0$    &           &               &85         &$41.5$         &           &       \\
HD 47054    &B8 V e         &24         &$-30.0$    &42         &$-31.5$        &           &               &           &       \\
HD 183914   &B8 V e         &17         &$16.8$     &37         &$20.8$         &           &               &           &       \\
HD 50138    &B8 V e         &27         &$-36.0$    &           &               &           &               &           &       \\
HD 58715    &B8 V var       &22         &$-21.5$    &44         &$-9.0$         &           &               &           &       \\
HD 23552    &B8 V ne        &           &           &33         &$24.7$         &           &               &           &       \\
HD 185037   &B8 V ne        &26         &$12.0$     &45         &$18.0$         &           &               &           &       \\
HD 199218   &B8 V nne       &17         &$4.7$      &35         &$4.2$          &           &               &95         &$7.7$\\
HD 91120    &B8/B9 IV/V     &23         &$-13.5$    &39         &$-15.0$        &           &               &           &       \\
HD 142926   &B9 pe          &18         &$-12.9$    &50         &$2.1$          &67         &$7.6$          &165        &$3.6$\\
HD 144      &B9 III e       &28         &$-30.6$    &53         &(DAC) $-188.6$ &80         &(DAC) $-187.1$ &178        &(DAC) $-129.6$\\
\end{longtable}

\begin{table}
  \caption{The same as in Table ~\ref{absrp}, but for the emission
component where it is present.} \label{emission}
  \begin{center}
    \begin{tabular}{llrr}
\hline
Star        &Spectral Type  & $V_{rote}$ &$V_{rade}$\\
\hline
HD 203374   &B0 IV pe    &20    &72\\
HD 45910    &B2 III e    &34    &81\\
HD 32991    &B2 V e      &57    &13\\
HD 164284   &B2 V e      &120   &-1\\
HD 148184   &B2 V ne     &50    &33\\
HD 45995    &B2 V nne    &46    &102\\
HD 65875    &B2.5 V e    &30    &53\\
HD 50820    &B3 IV e+... &59    &-87\\
HD 217891   &B6 V e      &50    &6\\
HD 192044   &B7 V e      &94    &7\\
HD 210129   &B7 V ne     &97    &-29\\
HD 47054    &B8 V e      &81    &0\\
HD 50138    &B8 V e      &53    &120\\
HD 199218   &B8 V nne    &131   &15\\
    \end{tabular}
  \end{center}
\end{table}

\begin{table}
  \caption{Values
of the confidence with which the accepted fit is better than the
fit with the one less component.}\label{ftest}
  \begin{center}
    \begin{tabular}{lccclccc}
\hline
Star & 4-3 comps & 3-2 comps & 2-1 comps & Star & 4-3 comps
& 3-2
comps & 2-1 comps\\
\hline
HD 5394     &           &0.9132     &           &HD 25940   &           &0.9972     &       \\
HD 53367    &1.0000     &           &           &HD 45725   &           &1.0000     &       \\
HD 203374   &0.9483     &           &           &HD 183362  &0.9807     &           &       \\
HD 206773   &           &1.0000     &           &HD 208057  &           &1.0000     &       \\
HD 200310   &           &           &0.9993     &HD 205637  &1.0000     &           &       \\
HD 212571   &           &0.9826     &           &HD 217543  &1.0000     &           &       \\
HD 44458    &0.9981     &           &           &HD 217050  &1.0000     &           &       \\
HD 200120   &           &           &1.0000     &HD 89884   &           &1.0000     &       \\
HD 193237   &1.0000     &           &           &HD 22192   &           &1.0000     &       \\
HD 45910    &0.1609     &           &           &HD 23302   &           &1.0000     &       \\
HD 37202    &1.0000     &           &           &HD 45542   &           &0.9998     &       \\
HD 36576    &           &0.9931     &           &HD 109387  &0.9996     &           &       \\
HD 212076   &           &1.0000     &           &HD 23480   &           &0.9998     &       \\
HD 32991    &           &0.6737     &           &HD 217891  &           &           &       \\
HD 58050    &           &0.9978     &           &HD 138749  &           &0.9936     &       \\
HD 164284   &           &0.7826     &           &HD 23630   &           &0.9986     &       \\
HD 41335    &1.0000     &           &           &HD 209409  &           &0.9954     &       \\
HD 52721    &           &0.9767     &           &HD 6811    &1.0000     &           &       \\
HD 58343    &           &0.9928     &           &HD 192044  &           &0.8070     &       \\
HD 148184   &           &0.9396     &           &HD 210129  &           &0.8927     &       \\
HD 202904   &           &0.9475     &           &HD 142983  &1.0000     &           &       \\
HD 65079    &1.0000     &           &           &HD 29866   &           &1.0000     &       \\
HD 28497    &1.0000     &           &           &HD 47054   &           &0.9086     &       \\
HD 45995    &0.1169     &           &           &HD 183914  &           &0.9975     &       \\
HD 10516    &0.9981     &           &           &HD 50138   &           &           &0.6758\\
HD 187567   &           &0.9747     &           &HD 58715   &           &0.9189     &       \\
HD 191610   &           &0.9990     &           &HD 23552   &           &           &1.0000\\
HD 32343    &           &0.7793     &           &HD 185037  &           &0.9500     &       \\
HD 65875    &           &1.0000     &           &HD 199218  &           &0.9693     &       \\
HD 60855    &0.9800     &           &           &HD 91120   &           &0.9778     &       \\
HD 37490    &1.0000     &           &           &HD 142926  &1.0000     &           &       \\
HD 50820    &           &0.6221     &           &HD 144     &1.0000     &           &       \\
    \end{tabular}
  \end{center}
\end{table}

\twocolumn

In order to be certain that we have accomplished the best fit, we
performed an F-test, between the fit that we accept as the best
and a fit with one component less. We present the results of the
F-test in Table ~\ref{ftest}, where we give the values of the
confidence with which the accepted fit is better than the fit with
the one component less. We did not perform an F-test between the
best fit and a fit with one component more, as this last one fit
presented extreme differences with the observed spectral line
profiles or because the values of the measured parameters go
against the classical theory for resonance lines \citep{dan03,
lyr04}.

\section{Results and discussion}

DACs phenomenon is quite common in O- and early B-type stars.
However, we found that DACs phenomenon is also observed in late
B-type stars (as in the case of HD 144). Moreover, many
researchers \citep{und75, mor77, mar78, dej79, doa82, sah84,
sah85, hut85}, even if they do not use the name DACs, observed the
same phenomenon. We point out also, that in \citet{und70, mar78,
dac80, doa82, sah84, sah85, hut85, doa91, cid98} only the known
DACs phenomenon was investigated. However, in the present paper
the similar phenomenon of SACs is introduced and we investigate
whether this phenomenon is able to explain the complex structure
of Mg II resonance lines in the stellar spectra of all the Be
spectral subtypes. Our result is that the SACs phenomenon is able
to explain this complex structure.

\begin{figure}
  \begin{center}
    \FigureFile(85mm,101mm){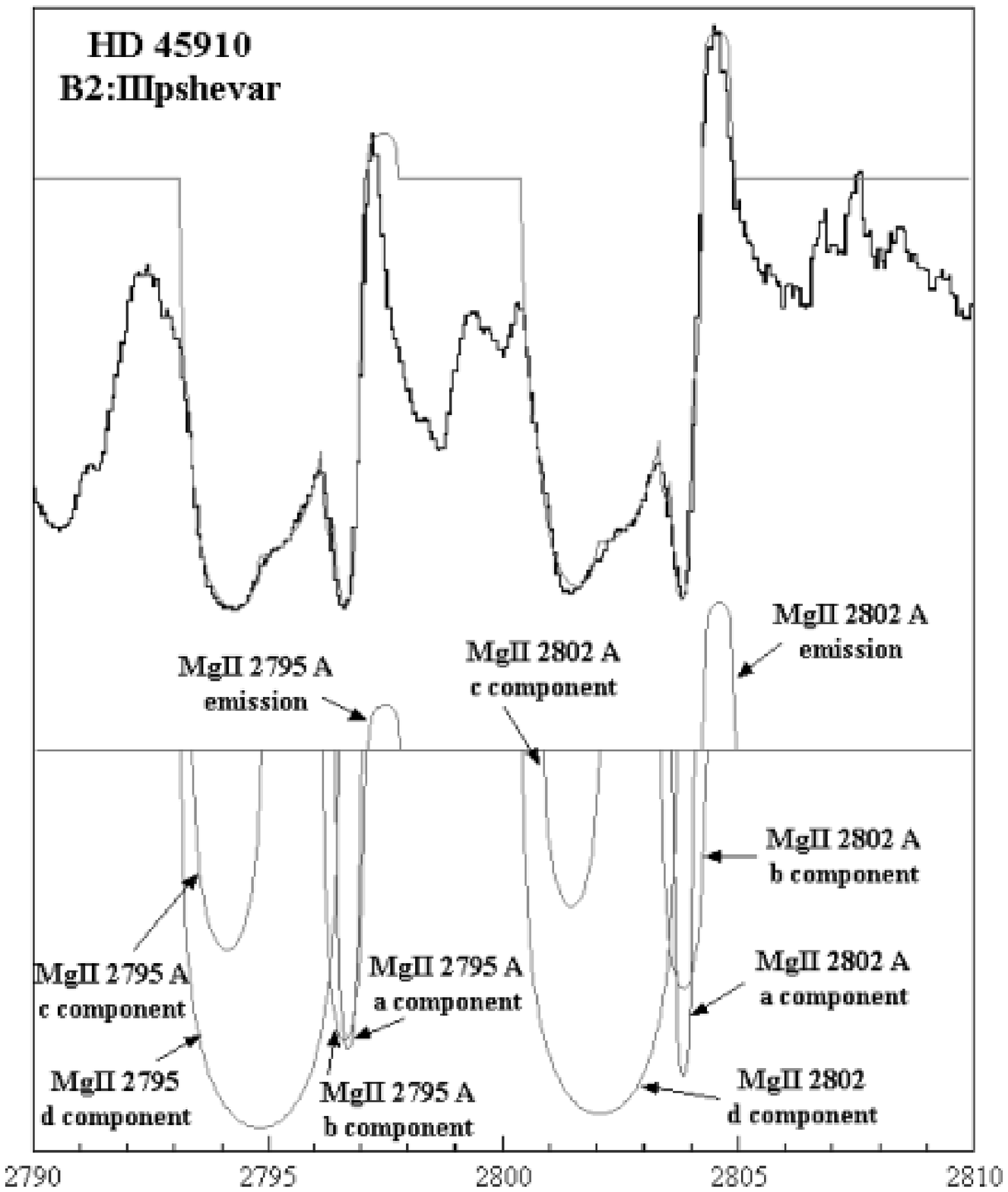}
    \FigureFile(85mm,101mm){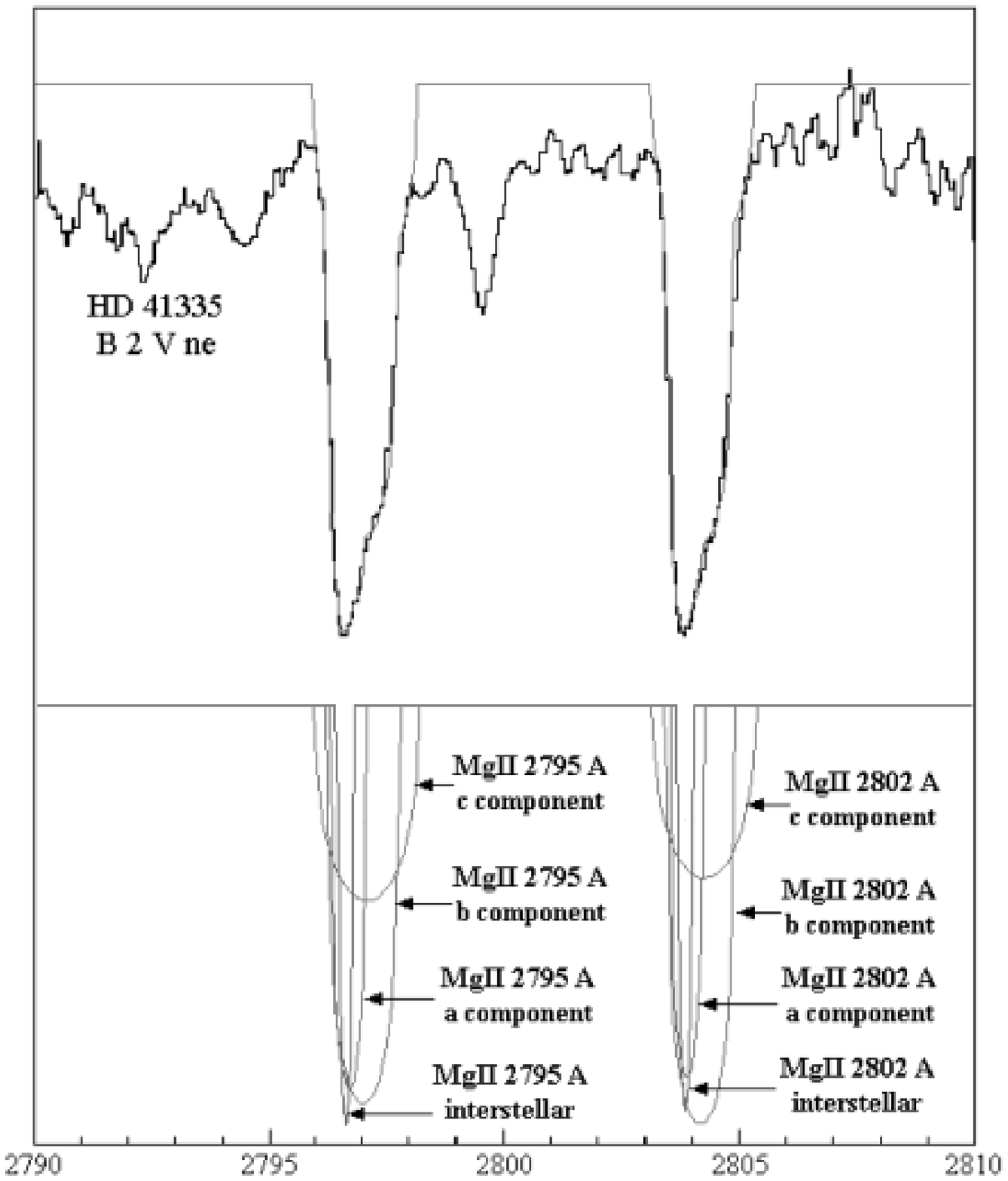}
  \end{center}
  \caption{ The Mg II resonance line profiles fitted with the model
for HD 45910 and HD 41335. The thick lines present observations,
and the thin lines the best fit. The DACs (HD 45910) and SACs (HD
41335) are present below.} \label{comp}
\end{figure}

\begin{figure}
  \begin{center}
    \FigureFile(85mm,33mm){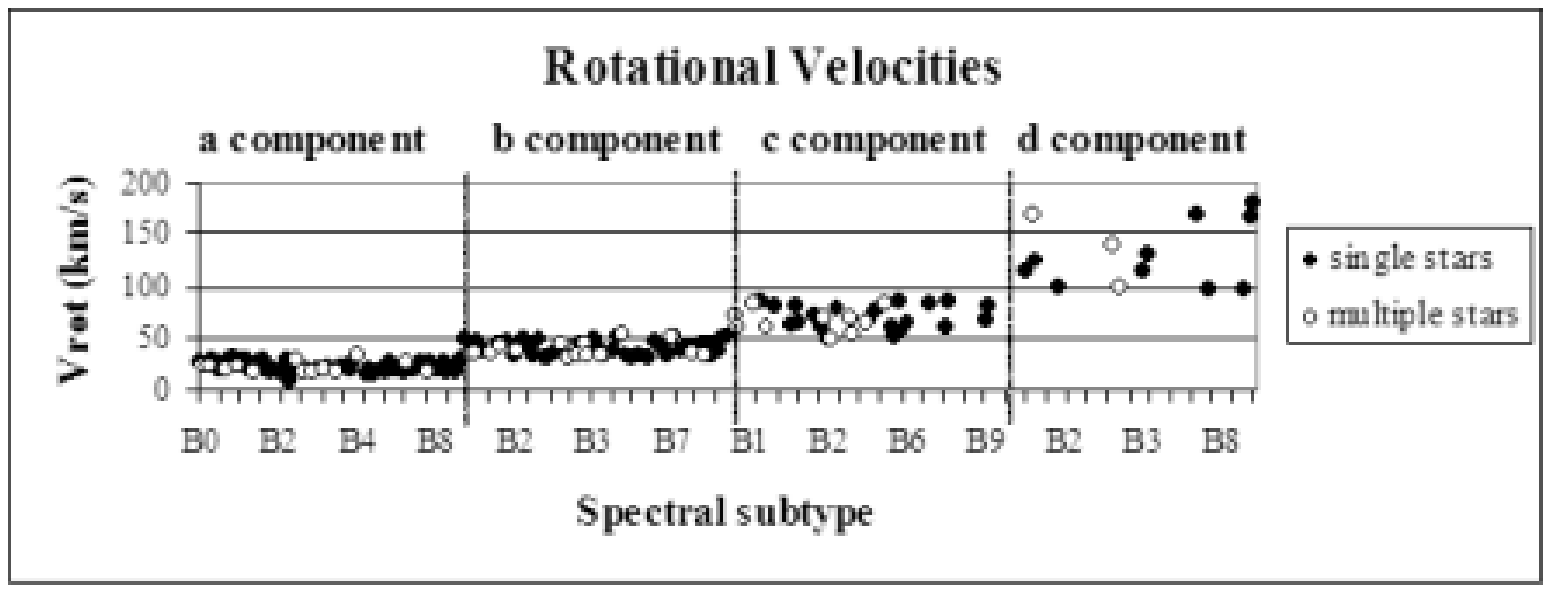}
  \end{center}
  \caption{Mean values for each spectral subtype of the rotational
velocities of all the SACs as a function of the spectral subtype.}
\label{Vrsep}
\end{figure}

\begin{figure}
  \begin{center}
    \FigureFile(85mm,33mm){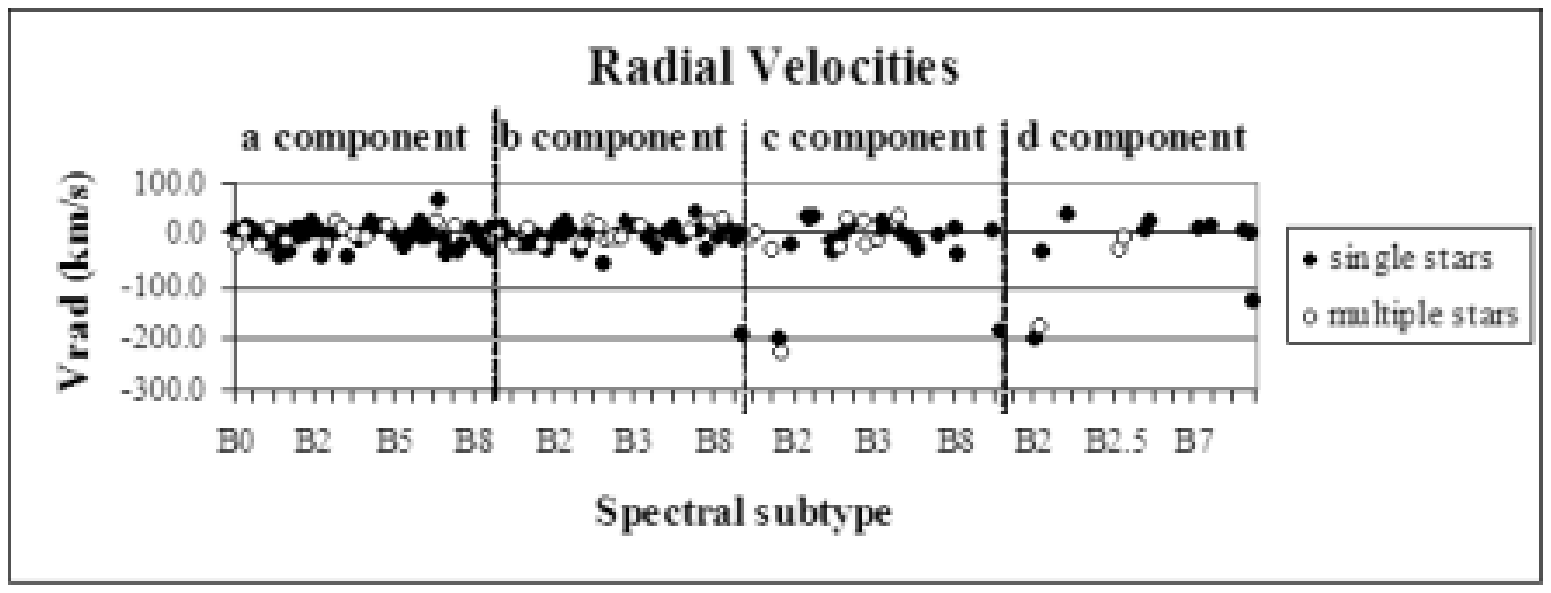}
  \end{center}
  \caption{Mean values for each spectral subtype of the radial
velocities of all the SACs as a function of the spectral subtype.}
\label{Vesep}
\end{figure}

\begin{figure}
  \begin{center}
    \FigureFile(85mm,33mm){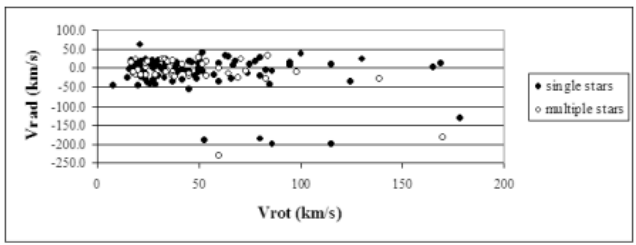}
  \end{center}
  \caption{Radial velocities of all the SACs as a function of the
respective rotational velocities.} \label{Ve-Vr}
\end{figure}

\begin{figure}
  \begin{center}
    \FigureFile(85mm,33mm){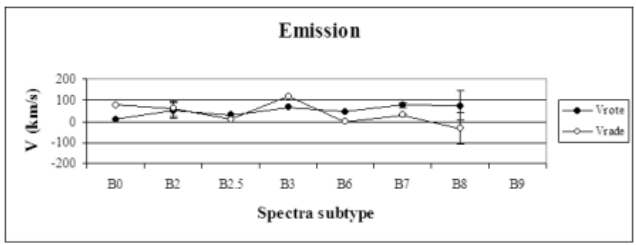}
  \end{center}
  \caption{Mean values for each spectral subtype of the rotational
and radial velocities of the emission component as a function of
the spectral subtype.} \label{em1}
\end{figure}

\begin{figure}
  \begin{center}
    \FigureFile(85mm,33mm){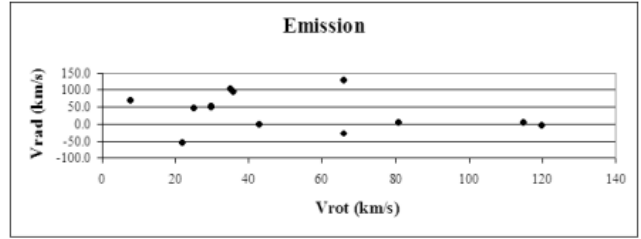}
  \end{center}
  \caption{Radial velocities of the emission component as a function
of the respective rotational velocities.} \label{em2}
\end{figure}

Using the model described above we find the best fit for the Mg II
resonance lines  of 64 Be stars given in Table ~\ref{data}. In
Fig. ~\ref{comp}, we give the best fit of the Mg II resonance
lines for two stars (HD 45910 and HD 41335). The first one (HD
45910) presents DACs, where the decomposition of the three
different components is easy, while the second one (HD 41335)
presents SACs, where, it is hard to decompose the three different
components, without the convenient model. So, one can see that the
model describes conveniently the Mg II complex profiles and the
complex structure of the regions where these lines are created. In
the studied IUE spectra, the interstellar lines are systematically
shifted to the red for +99 $\pm$ 16 $\rm {km\ s^{-1}}$. We used
the Hipparcos catalogues
\footnote{http://vizier.u-strasbg.fr/viz-bin/VizieR-3} and we
applied the corrections for the systemic velocity of individual
stars \citep{smi01} and for the orbital motion of the stars which
are members of binary systems. Our results, shown in Tables
~\ref{absrp} and ~\ref{emission} as well as in Figs. ~\ref{Vrsep}
- ~\ref{em2}, are correspondingly corrected.

We should note that in our sample 15 binary and multiple systems
are present (see Table ~\ref{data}). Considering that the origin
of DACs and SACs, in principle might be different than in single
stars, at the beginning we separately analyzed this kind of stars
(presented with open circles in Figs. ~\ref{Vrsep} -
~\ref{Ve-Vr}). However, we found that there is no systematic
difference between single and binary/multiple stars. Consequently,
in further analysis we will not treat binary/multiple systems in a
different way.

From the fit we obtained the rotational ($V_{rot_{i}}$) and radial
velocities ($V_{rad_{i}}$) for each Mg II resonance line
originating within the region.

In Tables ~\ref{absrp} and ~\ref{emission} we present the
kinematical parameters for the absorption Mg II resonance line
forming regions (Table ~\ref{absrp}), as well as for the emission
ones (Table ~\ref{emission}). As one can see, not all the studied
stars have an emission component, but only the ones listed in
Table ~\ref{emission}. One could assume that these emission
components are the emission part of P Cygni profiles formed by
scattering. In that case, their wavelength and width does not
represent the radial and rotational velocity. This could explain
the majority of positive radial velocities since in such a case
even in outflowing wind, lines will be widened and red shifted.
However, two cases with large negative radial velocities indicate
that the real picture may be more complicated. We know that the
disk models, in many cases, may produce theoretically only the
shape of line profiles, but they are not able to fit them. So, in
such cases, we may consider that the observed profile results from
a different mechanism. In Table ~\ref{absrp} we give the list of
stars, their spectral subtype (columns 1 and 2, respectively) and
the values of the rotational velocities (columns 3, 5, 7 and 9)
and the radial velocities (columns 4, 6, 8 and 10) of the
respective components. Also, in Table ~\ref{emission}, we present
the kinematical parameters of the emission components. In column 1
we present the list of stars, in column 2 their spectral subtype
and in columns 3 and 4 the values of the rotational and the radial
velocity, respectively. Let us point out, here, that the
calculated values correspond to the regions which create the SACs
or DACs. Especially, the obtained rotational velocities correspond
to the rotation of the region around itself and not around the
star.

In Figs.~\ref{Vrsep} and ~\ref{Vesep} we present, separately, the
rotational velocities and the radial velocities, respectively, of
all the SACs as a function of the spectral subtype. In Fig.
~\ref{Vesep} the radial velocity's values which correspond to the
DACs are clearly seen. In Fig.~\ref{Ve-Vr} we present the radial
velocities of all the SACs as a function of the respective
rotational velocities. In Fig.~\ref{em1} we present the rotational
and radial velocities' mean values of the two resonance lines of
the emission component as a function of the spectral subtype. Each
point refers to the mean value extracted for each spectral
subtype. Finally, the radial velocities of the emission component
as a function of the respective rotational velocities are given in
Fig.~\ref{em2}.

The values we have calculated lie within a small range and we can
obtain only the mean values of radial and rotational velocities
and their standard deviation. The points in the diagrams
correspond to the mean values of the velocities for each spectral
subtype and the error-bars which appear in some of the diagrams to
the standard deviation. This standard deviation is not only a
statistical error but it includes also the possible variation of
the inclination axis and the error of the spectral classification,
as the spectral classification which was based in the optical
range could not be appropriate for the UV range. This means that
the error-bars which appear in the diagrams include the
statistical error as well as the dispersion of the values due to
the different axis inclination.

The reproduction of the Mg II resonance lines $\lambda\lambda$
2795.523, 2802.698 ${\rm\AA}$ using the model where SACs are
present, suggests that the atmospherical regions, where the Mg II
doublet is created, may be described in a unique way for all the
studied Be stars. This result confirms the suggestion given by
\citet{dej79} that in Be stellar atmospheres exists a
concentration of low-ionization species in the stellar wind and it
is due to the occurrence of significant density variations. This
result is also in agreement with the results of \citet{mor77}, who
proposed that there are ``significant absorption features'' on the
left side of each resonance line, which are attributed to
``additional absorption within the stellar extended atmosphere''.
These ``significant absorption features'' can be the SACs that
appear in the spectra of the early type stars. \citet{dan03} and
\citet{lyr04} suggested that the peculiar phenomena observed in
the spectra of Oe and Be stars, such as the SACs, are due to
independent density regions in the stellar environment. Such
regions may be structures that cover all or a significant part of
the stellar disk (shells, blobs, puffs, bubbles) \citep{und75,
dej79, hen84, und84, bat86, gra87, lam80, wal92, cra96, riv97,
kap96, kap97, kap99, mar00}, interaction of fast and slow wind
components, CIRs, structures due to magnetic fields or spiral
streams as a result of the stellar rotation \citep{und84, mul84a,
mul84b, mul86, pri88, cra96, ful97, kap96, kap97, kap99, cra00}.
This is the common theory which explains the DACs phenomenon in
early type stars. We have found DACs phenomena in early B-type
stars (e.g HD 45910), as well as in late B-type stars (e.g. HD
144).

As one can see from Table 2, the SACs phenomenon appears to be a
classical one for the Be stars. In Fig.~\ref{comp} and in Table
~\ref{absrp} one can see that all the studied stars present
discernible (DACs) or indiscernible (SACs) components of the Mg II
resonance lines. The indiscernible components appear in the
spectra of all the stars of luminosity classes IV and V and most
of the stars of luminosity class III. It is very interesting that
the SACs are observed as discrete lines (DACs) in the spectra of
the three stars HD 193237 (B2 pe), HD 45910 (B2 III e) and HD 144
(B9 III e), because they present quite different radial shifts.
This means that the regions which create these lines move radially
with relatively large velocities producing lines shifted enough to
be easily observed in the spectra. On the other hand, in the case
of all the rest studied stars, the SACs of the Mg II resonance
lines present similar radial velocities, resulting to the SACs
being blended among themselves. In this case, we can distinguish
these lines by the systematic differentiations of the rotational
velocities.

The decomposition of the observed profiles for the Mg II regions
in Be stellar atmospheres confirms the existence of independent
density regions, since by using such a structure, we were able to
reproduce the resonance lines of Mg II in all the studied stars.
This decomposition is physically meaningful as it enables us to
detect kinematically different regions with different rotational
and radial velocities, as well as the optical depth and the column
density, for each of the Mg II independent density regions, which
produce DACs or SACs.

Our analysis shows that regions where the considered Mg II
resonance lines originate  ("blobs" and "pufs" created by  winds
or cool extended envelopes) may consist  of  more independent
density layers of matter with different kinematical properties
(one to four in the analyzed cases). We identified them by the
decomposition of observed Mg II lines in the number of components
which fit best the observed profile. Namely, depending on
particular star, we obtain that SACs or DACs may be divided in
several rotational velocity groups (average values for rotational
velocity groups found to be present at 64 considered stars are
22$\pm$5 $\rm {km\ s^{-1}}$, 41$\pm$7 $\rm {km\ s^{-1}}$,
69$\pm$11 $\rm {km\ s^{-1}}$ and 130$\pm$31 $\rm {km\ s^{-1}}$).
The corresponding radial velocities are near zero (-3.3$\pm$20.3
$\rm {km\ s^{-1}}$ - for the first density region; -3.6$\pm$20.6
$\rm {km\ s^{-1}}$ - for the second; -1.0$\pm$21.8 $\rm {km\
s^{-1}}$ - for the third one and +4.0$\pm$22.7 $\rm {km\ s^{-1}}$
for the fourth one). In the spectra of the stars HD 193237 (B2
pe), HD 45910 (B2 III e) and HD 144 (B9 III e) the SACs appear as
discrete components (DACs) and the radial velocities of the third
and the fourth density region are -205.7$\pm$21.7 $\rm {km\
s^{-1}}$ and -170.6$\pm$36.6 $\rm {km\ s^{-1}}$, respectively. In
the case of the star HD 144, the second component appears also as
a discrete component. The radial velocity of the region, which
creates this component is -188.6 $\rm {km\ s^{-1}}$ (Figs.
~\ref{Vrsep}, ~\ref{Vesep} and ~\ref{Ve-Vr}). The observed
velocity dispersion may be due to the different values of the
rotational axis inclination of the regions where the SACs are
created. The results presented above confirm that the Mg II
doublet is more or less stable for a given spectral type as
\citet{gur75} suggested. We did not find any variation of the
velocities in the Mg II regions with the luminosity class, except
in the case of the peculiar stellar spectra, where the SACs appear
as discrete lines (DACs), while \citet{kon76} proposed that, apart
from the difference among spectral subtypes, there is probably
difference among luminosity classes too.

We assume that independent density regions corresponding to
particular components of considered Mg II lines (one to four
components corresponding to one to four regions in the case of 64
stars analyzed here), lie all in the cool stellar envelope.
Depending on the temperature, different ions with different
Ionization Potential are created in different regions at different
distances from the star. This means that the spectral lines
observed in the spectra of Be stars derive from specific
atmospherical regions, different among themselves. The ions which
are created very close to the star lie in regions which present
spherical symmetry around the star (case A in Fig. 1). On the
other hand, the ions which are created at long distance from the
star lie in regions which present spherical symmetry around their
own center and not around the star (case B in Fig. 1). As the
ionization potential of the Mg II ions is I.P.=7.646 eV, the Mg II
ions can be created only at great distance from the center of the
star, meaning in the disk, where spherical symmetry around the
star cannot be accepted. This means that the Mg II ions lie at
regions which present topical or apparent spherical symmetry (case
B in Fig. 1). As a result, the Mg II spectral lines and their
SACs/DACs may derive only from such regions (case B in Fig. 1) and
not from classically spherical regions around the star (case A in
Fig. 1). This kind of density regions (blobs) have been proposed
by many researchers \citep{und75, dej79, lam80, hen84, und84,
mul84a, mul84b, mul86, bat86, gra87, pri88, wal92, cra96, ful97,
riv97, kap96, kap97, kap99, mar00, cra00} and are detected in many
other cases, as in active stars (WR 104, see Fig. 2) and many
quasars, as we observe DACs/SACs in their UV spectra
\citep{dan06}. This means that the density regions are a common
phenomenon, observed in different levels. The fact that we found
one to four components is accidental. In principle, there could be
more or less. Although the number of components is different in
different stars, we may conclude that the Mg II resonance lines
forming regions present a complex structure.

Our proposition that SACs phenomenon is responsible for the
structure of Mg II lines  means that we expect theoretically that
the Mg II components have similar radial velocities, within the
range of statistical error ($\sim10$ km/s). The problem was how to
distinguish them. The common idea \citep{doa82} is that the radial
velocity of the kinematically independent regions is a function of
the distance from the rapidly rotating Be star. According to that,
our first thought was to distinguish these regions according to
their rotational velocities, which is confirmed by our
calculations. Namely, the regions which create the Mg II lines
have similar radial velocities and different rotational
velocities.

In our sample we cannot detect high radial velocities, whatever
method we use (the proposed model or any other classical method).
The radial velocities of all the SACs, in all the studied stars
are about 0 km/s and only in the case of the stars HD 193237 (B2
pe), HD 45910 (B2 III e) and HD 144 (B9 III e), where we observe
DACs, we calculated radial velocities with values between -130
km/s and -230 km/s. We can detect the same phenomenon in the case
of Si IV in a sample of 68 Be stars (values of radial velocities
between -116 km/s and +25 km/s) \citep{lyr06}, as well as in the
case of Ha in 120 Be stars (values of radial velocities around 0
km/s) \citep{lyr05}. These results indicate that in the
atmospherical layers from the photosphere (very broad components
of Ha) to the cool envelope (Mg II resonance lines) we cannot
detect very high radial velocities. However, as we expected, the
presence of DACs in three stars of our sample indicate that from
the regions near to the star towards the ones away from the star,
the radial velocity increases, but it does not reach high values,
as happens in the case of Oe stars. In the case of specific Be
stars that present DACs with high radial velocities in some of
their spectra, e.g. 59 Cyg and $\gamma$ Cas \citep{doa89, tel94},
we should study each one of them as exception of the classical
rule. This proposition is based on the fact that the spectral
classification was made in the optical range and may not apply in
the UV spectral range \citep{wal71, wal84, wal87} (see also the
SIMBAD database: http://simbad.u-strasbg.fr/sim-fid.pl). This
means that the spectral classification in the optical and the UV
range are not always in accordance. As a result, some early Be
stars could be late Oe stars and we know that it is a common
phenomenon to observe high radial velocities of SACs/DACs in Oe
stars. However, in the case of Be stars, we do not observe the
same phenomenon. This difference in the behavior of density
regions in Oe and Be stars is very interesting and requires
further investigation. Finally, we should pay attention in the way
the radial velocities are calculated. The classical method
considered that the whole observed feature corresponds to only one
spectral line, meaning that the radial velocity was calculated by
the displacement of the deeper point of the observed feature.
Considering the SACs idea, the observed feature consists of a
number of spectral line components. As a result, the deeper point
of the observed feature is only the result of the synthesis of all
the SACs. In this case we should calculate the radial velocity of
every one of these components.

An emission component is present in the spectra of  B0, B2, B2.5,
B3, B6, B7, B8 and B9 type stars (Fig.~\ref{em1}). This means that
the emission does not appear in the spectra of the middle spectral
subtypes of the Be stars \citep{kon75}. The radial velocity of the
emission component decreases as the rotational velocity increases.
The emission component presents positive or negative radial
velocities. If one takes into account that negative radial
velocities exist, which is not in agreement with the assumption
that all emission components are the emission part of P Cygni
profiles formed by scattering, one can assume the following. The
obtained velocities correspond to the regions where the emission
component is created as e.g. strings, blobs, puffs, bubbles. This
means that the emission region may approach or move away from the
observer and its different position and motion around the star is
responsible of whether this value is positive or negative. In
Fig.~\ref{em1} and ~\ref{em2} one can see that, as the rotational
velocity increases, the radial velocity decreases, in contrast
with the relation of the two velocities of the absorption
components. A problem with the emission component is that it is
blended with absorption lines of other ions and thus it is
difficult to evaluate the rotational and radial velocities. As a
result the calculated values present greater statistical error
than in the case of the absorption components.

\begin{figure}
  \begin{center}
    \FigureFile(85mm,65mm){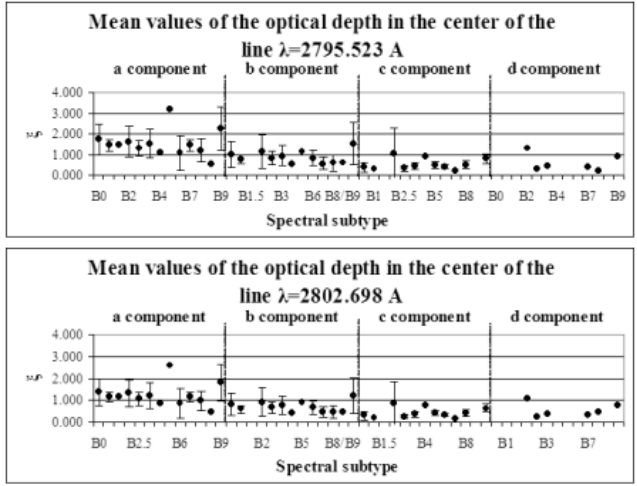}
  \end{center}
  \caption{Mean values of the optical depth $\xi$ in the center of
the line for all kinematically separated components of each
resonance line, as a function of the spectral subtype.} \label{xi}
\end{figure}

According to criteria described in \citet{dan03}, the same
components of the two resonance lines should have the same values
of rotational and radial velocities and the ratio of the optical
depth ($\xi$) should be the same as the respective ratio of
relative intensities. In Fig. ~\ref{xi} we present the mean values
of the optical depth ($\xi$) in the center of the line for all
kinematically separated components of each resonance line, as a
function of the spectral subtype. As one can see, the value of
$\xi$ of the first kinematical region is obviously higher than in
the rest three. It tends to decrease from the first to the fourth
kinematical region. That, also, may indicate physically separated
regions.

\section{Conclusions}

We applied the method developed in \citet{dan03} and \citet{lyr04}
on the Mg II resonance lines of 64 Be stars in order to
investigate the kinematical properties of the Mg II resonance
lines forming region. We obtained the rotational and radial
velocities which allow us to extract some general physical
properties for the Mg II regions of Be stars. Some interesting
results inferred from the investigations are the following: (i)
the proposed rotation model gives satisfactory results for the Mg
II $\lambda\lambda$ 2795.523, 2802.698 \AA\ resonance lines
region; (ii) the absorption atmospherical region where the Mg II
resonance lines are created presents a complex structure. It tends
to be composed by more than one kinematically independent regions
(Only four stars present simple structure). We found that the
kinematically independent regions rotate with different
velocities: 22 $\rm {km\ s^{-1}}$, 41 $\rm {km\ s^{-1}}$, 69 $\rm
{km\ s^{-1}}$, 130 $\rm {km\ s^{-1}}$. The respective radial
velocities are near zero for all these regions. These calculated
values lead us to accept that the Mg II resonance lines of the Be
stellar spectra present Satellite Absorption Components. (iii) the
rotational velocities of the found independent regions present a
uniform fluctuation with the spectral subtype. (iv) the emission
lines were detected in the earliest and latest spectral subtypes
with the majority of positive radial velocities with several
exceptions, i.e. with radial velocities within the range 0 to -87
$\rm {km\ s^{-1}}$, and taking into account that the positive
values, if lines are formed by scattering, are not radial
velocities.

\section{Acknowledgments}

This research project is progressing at the University of Athens,
Department of Astrophysics, Astronomy and Mechanics, under the
financial support of the Special Account for Research Grants,
which we thank very much. This work also was supported by Ministry
of Science and Environment Protection of Serbia, through the
projects Influence of collisional processes on astrophysical
plasma line shapes and Astrophysical spectroscopy of extragalactic
objects. Finally, we thank the anonymous referee for his/her very
useful comments.

\onecolumn
\appendix
\section{Calculation of the distribution functions L}

As we know the distribution function ($L$) of the absorption
coefficient ($k_{\lambda}$) has the same form as the distribution
function of each component of the spectral line. This means that
we can replace the distribution function of the absorption
coefficient $L_{i}$, with another expression of the distribution
function of each component.

It is also known that Be and Oe stars are rapid rotators. This
means that we accept that the main reason of the line broadening
is the rotation of the regions that produce each satellite
component of the whole observed spectral feature. These rapidly
rotating density regions may also present radial motion. For these
two reasons we search another expression for the distribution
function of the spectral line's components that has as parameters
the rotational and radial velocities of the spherical density
regions.

For a spherical density region, we assume the following
hypotheses: i) The natural broadening of the spectral lines
follows the Lorentz distribution; ii)Lamberts' sinus low stands
for each point of the spherical region; iii) The angular velocity
of rotation is constant.

In order to calculate the total radiation, we divide the spherical
layer in very thin cylindrical surfaces which are perpendicular to
the rotational axis. Lambert's low allows us to consider that the
luminosity from each point on the sphere is the same.

On the above cylindrical surfaces we also consider the surface
$dS$. According to Lambert's low, when this surface rotates with
an angular velocity $\omega$, its radiation intensity is:

\begin{equation}
\label{eq1} dI(\omega ) = Q(\omega )dS\cos \theta,
\end{equation}

\noindent where $\theta $ is the angle between the vertical on
$dS$ and the line of sight and

\[
Q(\omega ) = C_{1} {\frac{{\gamma}} {{(\omega - \omega _{k} )^{2}
+ \left( {{\frac{{\gamma}} {{2}}}} \right)^{2}}}},
\]
$C_{1}$ is a constant and $\gamma$ is the Lorentzian full width at
half maximum, which, in the case of the natural broadening, has
the value $\gamma \cong 10^{8} Hz$.

When the surface $dS$ does not rotate, the center of the formed
spectral line has the observed wavelength $\lambda_{0}$, which
corresponds to a frequency $\nu_{0}$. Thus, we have:

\[
\omega _{0} = 2\pi \nu _{0} = 2\pi {\frac{{c}}{{\lambda _{0}}} }.
\]

When the surface $dS$ rotates with a rotational velocity
$V_{rot}$, the center of the formed line has the wavelength
$\lambda_{k}$ and in this case we have $\omega _{k} = \omega _{0}
(1 - z\sin \varphi )$, where $z = {\frac{{V_{rot} }}{{c}}}$.

\begin{figure}[h]
  \begin{center}
    \FigureFile(85mm,76mm){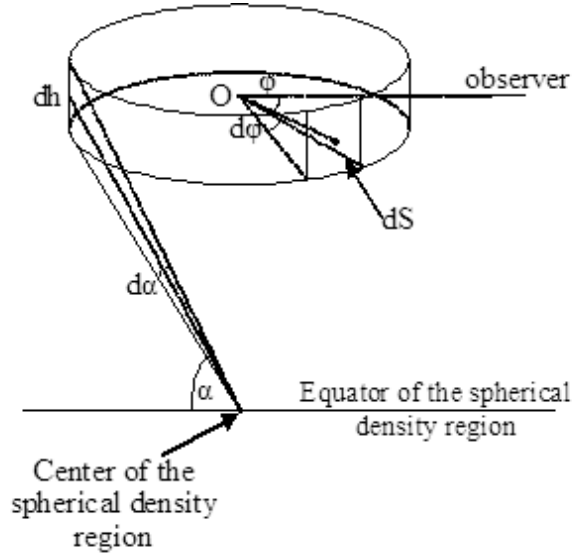}
  \end{center}
  \caption{Elementary ring of the spherical density region.}
\label{ring}
\end{figure}

We also have $\cos \theta \cong \cos \alpha \cos \varphi $. The
angles $\alpha$ and $\varphi$ are shown in Fig. ~\ref{ring}.

The surface $dS$ can be written as $dS = rdhd\varphi$, where $r$
is the radius of the cylinder, $d\varphi$ is the angle under which
the observer sees $dS$ and $dh$ is the height of $dS$.

Making the above substitutions in Eq. (\ref{eq1}), we have:

\[
dI(\omega ) = {\frac{{C_{1} rdh\gamma \cos \alpha \cos \varphi
d\varphi }}{{[\omega - \omega _{0} (1 - z\sin \varphi )]^{2} +
\left( {{\frac{{\gamma }}{{2}}}} \right)^{2}}}}
\]

Thus, the radiation intensity from the semi cylinder is

\[
I(\omega ) = {\int\limits_{{{ - \pi}  \mathord{\left/ {\vphantom
{{ - \pi} {2}}} \right. \kern-\nulldelimiterspace} {2}}}^{{{\pi}
\mathord{\left/ {\vphantom {{\pi}  {2}}} \right.
\kern-\nulldelimiterspace} {2}}} {{\frac{{C_{1} rdh\gamma \cos
\alpha \cos \varphi d\varphi}} {{[\omega - \omega _{0} (1 - z\sin
\varphi )]^{2} + \left( {{\frac{{\gamma}} {{2}}}} \right)^{2}}}}}}
\]

\noindent or

\[
I(\omega ) = {\frac{{4C_{1} rdh}}{{\gamma}} }{\int\limits_{{{ -
\pi} \mathord{\left/ {\vphantom {{ - \pi}  {2}}} \right.
\kern-\nulldelimiterspace} {2}}}^{{{\pi}  \mathord{\left/
{\vphantom {{\pi} {2}}} \right. \kern-\nulldelimiterspace} {2}}}
{{\frac{{\cos \alpha d\sin \varphi}} {{{\left[ {\left(
{{\frac{{\omega}} {{{\raise0.7ex\hbox{${\gamma }$}
\!\mathord{\left/ {\vphantom {{\gamma}
{2}}}\right.\kern-\nulldelimiterspace}\!\lower0.7ex\hbox{${2}$}}}}}}
\right) - \left( {{\frac{{\omega _{0}}}
{{{\raise0.7ex\hbox{${\gamma} $} \!\mathord{\left/ {\vphantom
{{\gamma}
{2}}}\right.\kern-\nulldelimiterspace}\!\lower0.7ex\hbox{${2}$}}}}}}
\right)\left( {1 - z\sin \varphi}  \right)} \right]}^{2} + 1}}}}}
\]

If we take that $\tilde {\omega}  = {\frac{{\omega}}
{{{\raise0.7ex\hbox{${\gamma} $} \!\mathord{\left/ {\vphantom
{{\gamma}
{2}}}\right.\kern-\nulldelimiterspace}\!\lower0.7ex\hbox{${2}$}}}}}$,
$\tilde {\omega} _{0} = {\frac{{\omega _{0}}}
{{{\raise0.7ex\hbox{${\gamma }$} \!\mathord{\left/ {\vphantom
{{\gamma}
{2}}}\right.\kern-\nulldelimiterspace}\!\lower0.7ex\hbox{${2}$}}}}}$,
$x = \sin \varphi $, we have

\[
I(\tilde {\omega} ) = {\frac{{4C_{1} r\cos \alpha dh}}{{\gamma
}}}{\int\limits_{ - 1}^{1} {{\frac{{dx}}{{[\tilde {\omega}  -
\tilde {\omega }_{0} (1 - zx)]^{2} + 1}}}}}
\]

Taking that $y = \tilde {\omega}  - \tilde {\omega} _{0} (1 -
zx)$, the above integral becomes

\[
I(\tilde {\omega} ) = {\frac{{4C_{1} r\cos adh}}{{\gamma \tilde
{\omega }_{0} z}}}{\int\limits_{\tilde {\omega}  - \tilde {\omega}
_{0} (1 + z)}^{\tilde {\omega}  - \tilde {\omega} _{0} (1 - z)}
{{\frac{{dy}}{{y^{2} + 1}}}}} .
\]

Finally, we have

\begin{equation}
\label{eq2} I(\tilde {\omega} ) = \left( {{\frac{{4C_{1} r\cos
\alpha dh}}{{\gamma}} }} \right)\left( {{\frac{{\arctan [\tilde
{\omega}  - \tilde {\omega} _{0} (1 - z)] - \arctan [\tilde
{\omega}  - \tilde {\omega} _{0} (1 + z)]}}{{\tilde {\omega} _{0}
z}}}} \right).
\end{equation}

The above function describes the radiation intensity from a visual
semi cylinder with radius r and height $dh$.

Under the angle $d\alpha$ is seen $dh$ from the center of the
spherical region. This cylinder rotates with a rotational velocity
$z = {\frac{{V_{rot} }}{{c}}}$ and a constant angular velocity
$\tilde {\omega} $.

We consider the function:

\[
P(\tilde {\omega} ,z) = {\frac{{\arctan [\tilde {\omega}  - \tilde
{\omega }_{0} (1 - z)] - \arctan [\tilde {\omega}  - \tilde
{\omega} _{0} (1 + z)]}}{{\tilde {\omega} _{0} z}}}
\]

We study the limit of this function in the case that the density
layer does not rotate, i.e. when $z \to 0$. In such a case:

\[
{\mathop {\lim} \limits_{z \to 0}} P(\tilde {\omega} ,z) =
{\mathop {\lim }\limits_{z \to 0}} {\frac{{\arctan [\tilde
{\omega}  - \tilde {\omega} _{0} (1 - z)] - \arctan [\tilde
{\omega}  - \tilde {\omega} _{0} (1 + z)]}}{{\tilde {\omega} _{0}
z}}}.
\]

We apply the De L' Hospital's low and we have:

\[
{\mathop {\lim} \limits_{z \to 0}} P(\tilde {\omega} ,z) =
{\mathop {\lim }\limits_{z \to 0}}
{\frac{{{\frac{{d}}{{dz}}}{\left\{ {\arctan [\tilde {\omega}  -
\tilde {\omega} _{0} (1 - z)]} \right\}} -
{\frac{{d}}{{dz}}}{\left\{ {\arctan [\tilde {\omega}  - \tilde
{\omega} _{0} (1 + z)]} \right\}}}}{{{\frac{{d}}{{dz}}}(\tilde
{\omega} _{0} z)}}}\]

\[{\mathop {\lim} \limits_{z \to 0}} P(\tilde {\omega} ,z) =
{\mathop {\lim }\limits_{z \to 0}} {\left[ {{\frac{{1}}{{1 +
[\tilde {\omega}  - \tilde {\omega} _{0} (1 - z)]^{2}}}} +
{\frac{{1}}{{1 + [\tilde {\omega}  - \tilde {\omega} _{0} (1 +
z)]^{2}}}}} \right]} = {\frac{{2}}{{(\tilde {\omega}  - \tilde
{\omega} _{0} )^{2} + 1}}}.
\]

It is obvious that in the non-rotating case this form corresponds
to the Lorentz's distribution for the naturally broadened spectral
lines.

\begin{figure}[h]
  \begin{center}
    \FigureFile(85mm,36mm){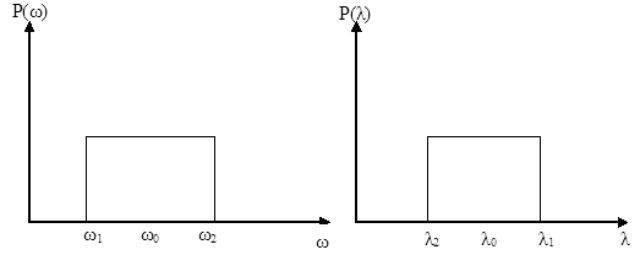}
  \end{center}
  \caption{Quadratic pulsation.} \label{puls}
\end{figure}

In the case when the rotation broadening $\vert
\lambda_{1}-\lambda_{2}\vert $ (or $\vert
\omega_{1}-\omega_{2}\vert$) is much larger than the natural
broadening (the natural broadening of a spectral line is of an
order of $10^{ - 3}$ - $10^{ - 4} \rm\AA$), the above function
$P(\tilde {\omega },z)$ presents the form of one quadratic
pulsation (see Fig. ~\ref{puls})

For each $\omega$ with $\omega _{1} < \omega < \omega _{2} $ the
relative shift is $z = {\frac{{{\left| {\Delta \omega}
\right|}}}{{\omega _{0}}} }$. For the point $\omega _{1} $ we have
${\frac{{\omega _{0} - \omega _{1}}} {{\omega _{0}}} } = z$ so
that $\omega _{1} = \omega _{0} (1 - z)$. Likewise, for the point
$\omega _{2} $ we have $\omega _{2} = \omega _{0} (1 + z)$. But
$\omega = 2\pi \nu = {\frac{{2\pi c}}{{\lambda}} }$. Thus,
$\lambda _{1} = {\frac{{\lambda _{0}}} {{1 - z}}}$ and $\lambda
_{2} = {\frac{{\lambda _{0} }}{{1 + z}}}$ with $\lambda _{1} >
\lambda _{2} $.

This means that $\Delta \lambda _{total} \equiv \lambda _{1} -
\lambda _{2} = {\frac{{2\lambda _{0} z}}{{1 - z^{2}}}}$ and so

 $\lambda _{\min}  \equiv \lambda _{2} = \lambda _{0} - {\frac{{\Delta \lambda
_{total}}} {{2}}} = \lambda _{0} - \lambda _{0} {\frac{{z}}{{1 -
z^{2}}}}$ and $\lambda _{\max}  \equiv \lambda _{1} = \lambda _{0}
+ {\frac{{\Delta \lambda _{total}}} {{2}}} = \lambda _{0} +
\lambda _{0} {\frac{{z}}{{1 - z^{2}}}}$.

We set $\rho = \lambda _{0} {\frac{{z}}{{1 - z^{2}}}}$ and
normalize to 1. In this way the function $P(\tilde {\omega} ,z)$
could be approximated with the function $f(\lambda )$ where:

\[
f(\lambda ) = {\left\{ {\begin{array}{l}
 {1\quad \quad \;{\left| {\lambda - \lambda _{0}}  \right|} < \rho}  \\
 {0\quad \quad {\rm otherwise}} \\
 \end{array}} \right\}}
\]

\begin{figure}[h]
  \begin{center}
    \FigureFile(85mm,44mm){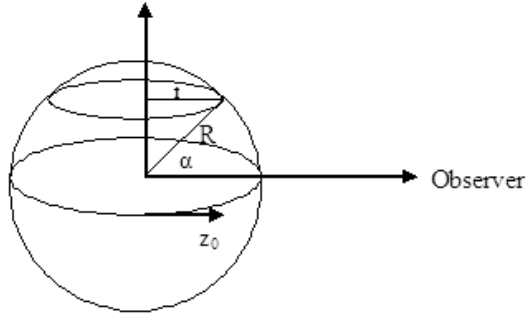}
  \end{center}
  \caption{Rotating spherical density region that produces SACs or
DACs.} \label{sphre}
\end{figure}

Now, we assume that the spherical density region rotates with
equatorial velocity $z_{0} = {\frac{{V_{0}}} {{c}}}$ (Fig.
~\ref{sphre}).

The points of the circle with radius r rotate with a velocity
$V_{rot} $ and for $r = R\cos \alpha $ we take

$V_{rot} = \omega r = \omega R\cos \alpha$.

We set $V_{0} = \omega R$. Also $\omega = const$. So: $V_{rot} =
{\frac{{V_{0}}} {{R}}}R\cos \alpha = V_{0} \cos \alpha $ and $z =
z_{0} \cos \alpha $. We have also $dh = Rd\alpha $ and Eq.
(\ref{eq1}) becomes

\[
dI(\tilde {\omega} ) = {\frac{{4C_{1} R}}{{\gamma}} } \cdot
{\frac{{\arctan [\tilde {\omega}  - \tilde {\omega} _{0} (1 -
z_{0} \cos \alpha )] - \arctan [\tilde {\omega}  - \tilde {\omega}
_{0} (1 + z_{0} \cos \alpha )]}}{{\tilde {\omega} _{0} z_{0} \cos
\alpha}} }.
\]

The integral of this equation is

\begin{equation}
\label{eq3} I(\tilde {\omega} ) = {\frac{{4C_{1} R}}{{\gamma}}
}{\int\limits_{ - {{\pi} \mathord{\left/ {\vphantom {{\pi}  {2}}}
\right. \kern-\nulldelimiterspace} {2}}}^{{{\pi}  \mathord{\left/
{\vphantom {{\pi}  {2}}} \right. \kern-\nulldelimiterspace} {2}}}
{{\frac{{\arctan [\tilde {\omega}  - \tilde {\omega} _{0} (1 -
z_{0} \cos \alpha ) - \arctan [\tilde {\omega}  - \tilde {\omega}
_{0} (1 + z_{0} \cos \alpha )]}}{{\tilde {\omega} _{0} z_{0} \cos
\alpha}} }}} \cos \alpha d\alpha.
\end{equation}

When we take into account the function $P(\tilde {\omega} ,z)$ the
above function becomes

\[
I(\tilde {\omega} ) = {\frac{{4C_{1} R}}{{\gamma}} }{\int\limits_{
- {{\pi} \mathord{\left/ {\vphantom {{\pi}  {2}}} \right.
\kern-\nulldelimiterspace} {2}}}^{{{\pi}  \mathord{\left/
{\vphantom {{\pi}  {2}}} \right. \kern-\nulldelimiterspace} {2}}}
{P(\tilde {\omega} ,z_{0} \cos \alpha )\cos \alpha d\alpha}}  .
\]

We approximate $P(\tilde {\omega}, z_{0} \cos \alpha )$ with
$f(\lambda )$ for and we take the integral for the observation
angle $\theta \in [ - \theta _{0} ,\theta _{0} ]$ from the
equatorial plane.

So we have that

\[
I(\tilde {\omega} ) \cong I_{1} = {\frac{{4C_{1} R}}{{\gamma}}
}{\int\limits_{ - \theta _{0}} ^{\theta _{0}}  {1 \cdot \cos
\theta d\theta}}.
\]

\[
I_{1} = {\frac{{4C_{1} R}}{{\gamma}}}{\int\limits_{ - \theta _{0}}
^{\theta _{0}}  {\cos \theta d\theta}} = {\frac{{4C_{1}
R}}{{\gamma}}}{\left[ {\sin \theta} \right]}_{ - \theta _{0}}
^{\theta _{0}}  = {\frac{{8C_{1} R}}{{\gamma}}}\sin \theta _{0} .
\]

If we normalize to 1, we have

\[
I_{1} = \sin \theta _{0} = \sqrt {1 - \cos ^{2}\theta _{0}}  .
\]

For the angle $\theta _{0} $ we have

\[
{\left| {\lambda - \lambda _{0}}  \right|} \le \rho _{0} =
{\frac{{\lambda _{0} z_{0} \cos \theta _{0}}} {{1 - z_{0}^{2} \cos
^{2}\theta _{0}}} }.
\]

For a wavelength $\lambda $ or for a shift $\Delta \lambda =
{\left| {\lambda - \lambda _{0}}  \right|}$ from the center of the
spectral line, the absorbing (or emitting) regions are those with
angular distance $\theta $ from the equatorial plane, with
${\left| {\theta}  \right|} \le \theta _{0} $.

For the equatorial plane we have

\[
\Delta \lambda = {\frac{{\lambda _{0} z_{0} \cos \theta _{0}}} {{1
- z_{0}^{2} \cos ^{2}\theta _{0}}} }.
\]

From this equation we can calculate the angle $\theta _{0} $ as
follows:

\[
\cos \theta _{0} = {\frac{{ - \lambda _{0} \pm \sqrt {\lambda
_{0}^{2} + 4\Delta \lambda _{}^{2}}} } {{2\Delta \lambda z_{0}}}
}.
\]

As $\theta_{0}$ is between $-\pi/2$ and $\pi/2$ we have
$cos\theta_{0} \ge 0$ and so

\[
\cos \theta _{0} = {\frac{{ - \lambda _{0} + \sqrt {\lambda
_{0}^{2} + 4\Delta \lambda _{}^{2}}} } {{2\Delta \lambda z_{0}}}
}.
\]

Thus, the distribution function $I_{1}$ takes its final form:

$I_{1} = \sqrt {1 - \cos ^{2}\theta _{0}}  $ if $\cos \theta _{0}
= {\frac{{ - \lambda _{0} + \sqrt {\lambda _{0}^{2} + 4\Delta
\lambda _{}^{2}}} } {{2\Delta \lambda z_{0}}} } < 1$ and

$I_{1} = 0$, otherwise.

It is obvious that the distribution function $I_{1}$ is a function
of the wavelength ($\lambda$). This means that $I_{1} =
I_{1}(\lambda)$. This distribution function has the same form as
the distribution function of the absorption coefficient $L$ and
may replace it (in $e^{ - L\xi} $), when the main reason of the
line broadening is the rotation. We name it Rotation distribution
function.

The spectral line profile, which is formed by a spherical density
region, is reproduced by the function $e^{ - L\xi} $ by applying
the appropriate value of the rotational velocity $V_{rot} $ (from
$z_{0})$, the radial velocity $V_{rad}$ (from ${\frac{{V_{rad}}}
{{c}}} = {\frac{{\lambda _{0} - \lambda _{lab} }}{{\lambda
_{lab}}} })$ and the optical depth $\xi$ in the center of the
line.

\end{document}